\documentclass[runningheads]{llncs}
\usepackage[T1]{fontenc}
\usepackage{graphicx}
\usepackage{tikz}
\usepackage{placeins}
\usetikzlibrary{arrows.meta}

\usepackage{tikz}
\usepackage{pgfplots}
\pgfplotsset{compat=1.18} 

\usepackage{amsmath}
\usepackage{float}
\usepackage{booktabs}
\begin{document}
\title{LATS-RCA: Language Agent Tree Search for Root Cause Analysis in Microservices}
\titlerunning{LATS-RCA for Root Cause Analysis in Microservices}
%
%
\author{Alexander Naakka \inst{1}\orcidID{0009-0005-4487-7060} \and
Yuqing Wang\inst{2}\thanks{Corresponding author.}\orcidID{0000-0003-0175-005X} \and
Mika V Mäntylä\inst{2}\orcidID{0000-0002-2841-5879}}
\authorrunning{Naakka et al.}
%
\institute{Zoner Oy, Helsinki, Finland \\
\email{alexandru.constantinov@group.one}
\and
University of Helsinki, Helsinki, Finland \\
\email{firstname.lastname@helsinki.fi}
}
%
\maketitle              
\begin{abstract}
  Recent advances in large language models (LLMs) have enabled early attempts to automate root cause analysis (RCA) in microservice systems (MSS). However, existing approaches typically rely on a linear reasoning process that proceeds along a single diagnostic path. In this paper, we propose the Language Agent Tree Search for RCA (LATS-RCA) in MSS. LATS-RCA formulates RCA as a reflection-guided tree-structured search over root-cause hypotheses, where multiple agents iteratively analyze logs and metrics to collect evidence, and reflection scores guide the search toward the most likely root causes for abnormal services. We evaluate LATS-RCA on the open benchmark (LO2), achieving 91.3\% diagnostic accuracy and analyzing the associated computational cost. Variation among the frontier-tier LLMs (Claude Sonnet 4.5, GPT-5, and Gemini 3 Pro) is small, between 89.7\% and 91.3\%, demonstrating our approach is model-agnostic. We also conduct an exploratory study by evaluating LATS-RCA on real-world incidents from a web-hosting company's (Zoner Oy) production MSS that serves over 300,000 websites across Europe. We find that LATS-RCA correctly diagnoses 65.1\% of the production incidents on average over multiple runs. This reveals key challenges of real-world RCA, including multi-factor root causes, large-scale system complexity, and incomplete observability, which are absent from open benchmarks. Future work should develop more realistic open datasets for RCA and validate LATS-RCA with additional datasets. Our replication package is available at https://github.com/kottinov/lats-rca.

  
  

\end{abstract}
\section{Introduction}
Microservice systems (MSS) are widely adopted in production environments to support scalable, continuously evolving software platforms \cite{YuqingFSE2025}. In MSS, functionality is decomposed to independent services that may scale across regions, depend on external APIs and data platforms, and jointly underpin business processes. Production anomalies are common in microservice environments \cite{zhou2018fault}. 

Root cause analysis (RCA) plays a central role in  microservice operations by enabling engineers to identify the underlying causes of anomalies and to prevent future recurrence \cite{Ding2023TraceDiag,Adha2025}. Without an understanding of why failures occur, similar anomalies can reappear~\cite{2024surveyrootcause}. As a result, RCA is essential not only for short-term incident recovery but also for long-term system reliability. However, in practice, performing RCA on microservices remains largely a manual and experience-driven process due to the complexity of microservice operations \cite{zhou2018fault,Ding2023TraceDiag}. Engineers must reason over diverse operational signals, correlate symptoms, and interpret complex dependencies. Such manual RCA is labor-intensive, difficult to scale and its effectiveness depends on individual expertise \cite{zhou2018fault,Ding2023TraceDiag}.

Prior works primarily perform root cause localization by ranking candidate components without reasoning about the underlying causes. For example, they use service dependency graphs and causal inference to facilitate RCA with GNN and causal inference models \cite{Nezha2023,zheng2024mulan}.
With such approaches conducting further RCA requires manual effort to interpret and validate the ranked candidates \cite{YuqingFSE2025}. 

Recent advances in LLMs provide new opportunities for automating RCA in MSS. LLMs can potentially determine the root causes of anomalies through reasoning. Existing studies, RCAgent \cite{wang2023rcagent} and MicroRCA \cite{MicroRCA}, represent important initial steps in this direction. However, these approaches perform RCA along a linear reasoning process, in which a single sequence of analysis steps leads to a final result. However, in practical MSS environments, an observed anomaly may initially implicate multiple services or components \cite{RCALuan2024,ping2026}, making it difficult to determine the root cause through a purely linear reasoning process.


In this paper, we propose LATS-RCA that formulates RCA as a reflection-guided tree-structured search via the Language Agent Tree Search (LATS) algorithm \cite{LATSzhou24}. LATS combines Monte Carlo Tree Search (MCTS) with LLM reflection, modeling reasoning as an explicit tree exploration process in which multiple candidate reasoning paths are expanded, scored, and selectively pursued before reaching a final decision. LATS has demonstrated strong performance in multi-step reasoning settings, such as code generation, question answering, and interactive decision-making~\cite{LATSzhou24}.

LATS-RCA performs RCA for microservices based on their execution logs and performance metrics. 
Logs capture runtime events, whereas metrics summarize resource and performance behaviors \cite{YuqingSaner2025}. To analyze microservice behavior, LATS-RCA leverages multiple LLM-driven agents that iteratively reason over logs and metrics in a coordinated manner. Rather than treating RCA as a single linear reasoning process, 
LATS-RCA explicitly maintains multiple candidate root causes during search and incrementally gathers evidence from logs and metrics to evaluate how well each candidate explains the observed behavior. LLM-based reflection scores are used to evaluate intermediate diagnostic states and guide the search toward the most likely root cause. In the end, LATS-RCA outputs a root cause for each abnormal microservice. 




We evaluate LATS-RCA on Light-OAuth2 (LO2), an open-source industrial MSS, 
using the publicly available LO2 dataset \cite{L02paper}. LATS-RCA achieves a diagnostic accuracy of 91.3\%, outperforming the single-agent ReAct (39.8\%) baseline and multi-agent ReAct baseline (57.4\%).  Since no directly comparable LLM-based RCA approach exists under our experimental setting, we construct these two ReAct baselines ourselves as linear-reasoning references (detailed in Section 4.1). Each investigation incurs an average cost of 53.1 API calls, 156K tokens, and 9.1 minutes of execution time. The computational cost of LATS-RCA is primarily due to its systematic exploration of a larger hypothesis space, rather than increased evidence collection. 

Additionally, we conduct an exploratory study by evaluating LATS-RCA in Zoner Oy's real production MSS (Prod), which 
serves over 300,000 websites across Europe. Prod consists of independently developed services in a polyglot stack (Node.js, Rust, Python, and Go), generating heterogeneous logs and metrics. We collect 37 real-world incidents from Prod over a six-month period and construct a dataset using their corresponding logs and metrics. Each incident is manually analyzed to identify and verify its root cause(s), ensuring reliable ground-truth labels.
On Prod, LATS-RCA achieves an average diagnostic accuracy of 65.1\% (24.1 out of 37 incidents) across multiple runs, with results ranging from 62\% to 70\%, which is lower than 91.3\% observed on the LO2 benchmark. This Prod setting also incurs a higher reasoning cost, with each investigation involving an average of 75 API calls, 220K tokens, and 13 minutes of execution time. This Prod deployment provides empirical evidence of key challenges in production RCA that are absent from controlled benchmarks, including multi-factor failures, large-scale system complexity, and incomplete observability.

In summary, this paper makes these main contributions:
\begin{itemize}
    \item We propose LATS-RCA, an LLM-based multi-agent framework for RCA in MSS. LATS-RCA iteratively analyzes logs and metrics by formulating diagnosis as a reflection-guided tree search over competing root-cause hypotheses.
    \item We evaluate LATS-RCA on the LO2 benchmark, achieving 91.3\% diagnostic accuracy and providing a quantitative analysis of the associated computational costs.
    
    \item We conduct an exploratory study of LATS-RCA in a real production microservice environment, demonstrating its applicability beyond controlled benchmarks and providing insights into real-world RCA challenges. 
\end{itemize}

\section{Related work}

Many deep learning approaches to RCA in MSS rely on graph-based modeling and causal inference. For example, Nezha \cite{Nezha2023} transforms metrics, traces, and logs into a homogeneous event representation, constructs event graphs, and localizes root causes by comparing event patterns between fault-free and fault-affected phases. DiagFusion \cite{zhang2023diagfusion} unifies metrics, logs, and traces into event embeddings and applies GNNs over a service dependency graph to localize the faulty service instances. 
CausalRCA \cite{CausalRCA} learns a weighted causal graph over service metrics and applies PageRank to score and rank suspicious services. MULAN \cite{zheng2024mulan} learns causal structures from metrics and logs using contrastive multimodal representation learning and applies network propagation to localize root cause services. While these methods effectively narrow the investigation scope by localizing potentially faulty components, they do not identify the underlying root cause.

LLM-based multi-agent systems for RCA over multimodal monitoring data offer a promising direction to address this gap, though work in this area remains scarce given its novelty. To the best of our knowledge, RCAgent \cite{wang2023rcagent} and MicroRCA \cite{MicroRCA} are the only related works. RCAgent formulates RCA as an execution-based reasoning task over distributed traces, logs, and metrics. Instead of directly processing raw monitoring data, RCAgent generates executable analysis programs to query, aggregate, and preprocess multimodal monitoring data (i.e., Python code for filtering logs, aggregating performance metrics over time windows, or extracting statistics from traces), and reasons over the results returned by program executions. RCAgent outputs benchmark-specified diagnostic answers composed of structured fields (e.g., fault time span and faulty component), optionally accompanied by explanatory text derived from program execution results. MicroRCA adopts a pipeline-oriented approach that combines extensive modality-specific preprocessing with prompt-driven LLM reasoning. Given a fault time window, MicroRCA applies extensive modality-specific preprocessing, including log parsing and template-based filtering, trace anomaly detection using heuristic rules, and metric selection and filtering. The processed signals from logs, traces, and metrics are then summarized and provided to an LLM through carefully designed prompts, enabling the model to synthesize cross-modal evidence and infer the root cause. MicroRCA outputs structured RCA results in JSON form, including the identified fault component, the root-cause reason, and an accompanying reasoning trace.



\section{LATS-RCA Architecture}
\label{sec:LATS-RCA}
LATS-RCA consists of two diagnostic agents (log and metrics), coordinated by a supervisor. Each agent operates over a distinct observability modality: the log agent explores log data, and the metric agent analyzes metric data and generates comparative visualizations. The supervisor coordinates cross-modal handoff, sequences agent execution, and performs final correlation assessment. 

Each agent independently executes the diagnostic process on its modality. We model this process as a finite-horizon sequential decision process $\langle S, A, T, R \rangle$. A state $s \in S$ captures the current observations, active hypotheses, and investigation context, including the exploration trajectory, service dependencies, and search depth. An action $a \in A$ corresponds to invoking an analysis tool with specified arguments or generating a reasoning step, conditioned on the current investigation trajectory. The action space is instantiated with modality-specific tools: the log agent uses our custom tools for file listing, content retrieval, and pattern-based search, while the metric agent uses our custom tools for loading, comparing, and visualizing time-series data. The transition operator $T$ is realized through tool execution: given $(s,a)$, executing the selected tool(s) yields a successor state $s' = T(s,a)$. When a candidate action includes multiple tool invocations, these are executed in parallel, and each invocation produces a \texttt{ToolMessage} that records the tool's output and is appended to the trajectory. We refer to each \texttt{ToolMessage} returning monitoring data as an evidence item, and cap the number of evidence items at 10 per agent run. The reward signal $R$ is computed from a structured reflection of candidate next steps. This formulation casts RCA as a search problem over diagnostic states. 

Each agent performs an independent diagnostic search by maintaining a diagnostic search tree, which is
iteratively extended as the search proceeds using an LLM-reflection guided MCTS algorithm.
In this search tree, each node corresponds to a diagnostic state $s \in S$, and each edge
represents an investigative action $a \in A$. The search starts from a root node that
represents the initial diagnostic state. At each iteration, executing an investigative action produces a successor diagnostic state, which is added to the tree as a
new node. Each search iteration consists of four distinct phases: selection, expansion, reflection with reward computation, and backpropagation.

\begin{figure}[h]
    \centering
    \begin{tikzpicture}[
        scale=0.62,
        node/.style={circle, draw, thick, minimum size=0.80cm, fill=white, font=\scriptsize},
        visited/.style={circle, draw, thick, minimum size=0.80cm, fill=blue!15, font=\scriptsize},
        selected/.style={circle, draw, very thick, minimum size=0.80cm, fill=green!25, font=\scriptsize, draw=green!60!black},
        newnode/.style={circle, draw, thick, dashed, minimum size=0.80cm, fill=orange!15, font=\scriptsize, draw=orange!80},
        edge/.style={->, >=stealth, thick, black!70},
        selectedpath/.style={->, >=stealth, very thick, blue!70},
        newedge/.style={->, >=stealth, thick, dashed, orange!80},
        valuelabel/.style={font=\scriptsize, inner sep=1.5pt, fill=white, rounded corners=1pt},
        actionlabel/.style={font=\scriptsize\itshape, text=black!60, sloped, pos=0.4},
        >=stealth
    ]
        \node[node] (s0) at (0,0) {$s_0$};

        \node[visited] (s1) at (-2.5,-2.0) {$s_1$};
        \node[node] (s2) at (0,-2.0) {$s_2$};
        \node[node] (s3) at (2.5,-2.0) {$s_3$};

        \node[node] (s11) at (-3.8,-4.2) {$s_{11}$};
        \node[selected] (s12) at (-1.6,-4.2) {$s_{12}$};
        \node[node] (s21) at (0.4,-4.2) {$s_{21}$};

        \node[newnode] (s121) at (-2.4,-6.3) {$s_{121}$};
        \node[newnode] (s122) at (-0.8,-6.3) {$s_{122}$};

        \draw[selectedpath] (s0) -- node[actionlabel, above] {$a_1$} (s1);
        \draw[edge] (s0) -- node[actionlabel, above] {$a_2$} (s2);
        \draw[edge] (s0) -- node[actionlabel, above] {$a_3$} (s3);

        \draw[edge] (s1) -- node[actionlabel, above] {$a_{11}$} (s11);
        \draw[selectedpath] (s1) -- node[actionlabel, above] {$a_{12}$} (s12);
        \draw[edge] (s2) -- node[actionlabel, above] {$a_{21}$} (s21);

        \draw[newedge] (s12) -- node[actionlabel, above] {$a_{121}$} (s121);
        \draw[newedge] (s12) -- node[actionlabel, above] {$a_{122}$} (s122);

        \node[valuelabel] at (0,0.42) {$\mathrm{UCT}=0.68$};

        \node[valuelabel] at (-2.5,-1.4) {$\mathrm{UCT}=0.72$};
        \node[valuelabel] at (0,-1.4) {$\mathrm{UCT}=0.55$};
        \node[valuelabel] at (2.5,-1.4) {$\mathrm{UCT}=0.61$};

        \node[valuelabel] at (-3.8,-3.65) {$\mathrm{UCT}=0.58$};
        \node[valuelabel] at (-1.6,-3.65) {$\mathrm{UCT}=0.76$};
        \node[valuelabel] at (0.4,-3.65) {$\mathrm{UCT}=0.49$};

        \node[valuelabel] at (-2.4,-6.75) {$V=0.82$};
        \node[valuelabel] at (-0.8,-6.75) {$V=0.63$};

        \node[fill=white, inner sep=4pt, align=left, font=\scriptsize] at (4.55,-4.1) {
            \tikz{\node[circle, draw, thick, minimum size=0.35cm, fill=blue!15] {};}
            Sel. path \\[1pt]
            \tikz{\node[circle, draw, very thick, minimum size=0.35cm, fill=green!25, draw=green!60!black] {};}
            Expanded \\[1pt]
            \tikz{\node[circle, draw, thick, dashed, minimum size=0.35cm, fill=orange!15, draw=orange!80] {};}
            New \\[4pt]
        };

    \end{tikzpicture}
    \caption{Illustrative LATS search tree schematic.
    The nodes are annotated with the estimated value  $V_i$ and UCT scores computed in Eq.~\ref{eq:uct}.}
    \label{fig:lats-cycle}
\end{figure}
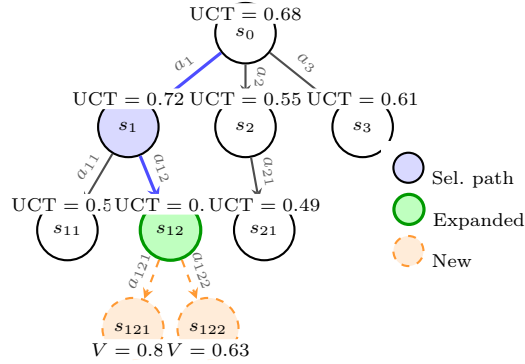

In the selection phase, the agent chooses a node (i.e., a diagnostic state $s_i$) from the current search tree for further expansion. Selection is guided by the Upper Confidence Bound for Trees (UCT) score: 
\begin{equation}
\text{UCT}(s_i) = V_i + c_{\text{uct}}\sqrt{\frac{\ln n_p}{n_i}},
\label{eq:uct}
\end{equation}
where $V_i$ denotes the value estimate for $s_i$, $n_p$ is the parent visit count, $n_i$ is the node visit count, and $c_{\text{uct}}$ is the exploration constant. Following the common UCT convention, we set $c_{\text{uct}} = 1.0$;
a sensitivity analysis with $c_{\text{uct}} \in \{0.5, 2.0\}$ resulted in accuracy
variations of less than 3\%. Equation~\ref{eq:uct} computes a selection score that trades off exploitation
(high $V_i$) and exploration (low $n_i$), and is used to prioritize promising
yet under-explored diagnostic hypotheses. Figure~\ref{fig:lats-cycle} illustrates an example of node selection and expansion under the UCT criterion.

In the expansion phase, at each expanded leaf node, the agent samples $N=5$ candidate actions using temperature-scaled decoding ($\tau=0.7$). We choose $\tau=0.7$ to balance determinism and exploration: lower temperatures reduce action diversity, whereas higher temperatures increase low-utility proposals. We set $N=5$ following prior LATS work \cite{LATSzhou24}; in pilot runs, this produced 2--3 distinct candidate actions per expansion while keeping API cost manageable.

In the reflection phase, each sampled candidate action is scored by an LLM reflection module along three axes: evidence quality~($e$), diagnostic completeness ($c_{\text{comp}}$), and internal consistency ($k$). For each axis, the LLM is provided with an explicit scoring criterion and required to return an integer score in the range 0–10 via structured output. Evidence quality~($e$) assesses whether the evidence items 
are relevant to the failure and sufficient to support the diagnosis, penalizing cases where critical evidence items are missing or irrelevant. Diagnostic completeness ($c_{\text{comp}}$) measures whether the response enumerates plausible root-cause hypotheses and systematically confirms or rules out each one. The score is reduced when obvious hypotheses are omitted or when the response considers only a single cause. Internal consistency ($k$) evaluates whether the claims are logically coherent, free of contradictions, and properly supported by the cited evidence. The three scores are aggregated into a scalar reflection score $r$ and normalized to [0.0, 1.0] by averaging and scaling:
\begin{equation}
 r = \frac{e + c_{\text{comp}} + k}{3\times10},
 \label{eq:reflection-score}
\end{equation}
where $r$ summarizes the quality of a proposed next action. 

To encourage robust decision-making under sampling variability, we compute a self-consistency term. An action signature is defined as the tuple (tool name, {argument key names}); two actions share a signature when they invoke the same tool with the same set of argument names, regardless of argument values. For candidates invoking multiple tools, the composite signature is defined as the sorted tuple of individual signatures, while candidates with no tool calls are assigned a sentinel signature. Let $\sigma$ denote the composite action signature and let $n_\sigma$ be its frequency among the $N$ sampled candidate actions; then $sc(\sigma)=n_\sigma/N$. The final reward associated with the corresponding state transition is defined as:
\begin{equation}
R = w\, r + (1-w)\, sc,
\label{eq:reward}
\end{equation}
where $w=0.5$ balances reflection quality and self-consistency. 
Equation~\ref{eq:reward}
thus favors actions that are both high-quality under reflection ($r$) and stable under
sampling ($sc$), reducing sensitivity to single-sample outliers.

In the backpropagation phase, the combined reward $R$ is propagated backward along the selected search path via online mean updating. Search iterations terminate when a candidate solution is confirmed, the depth 
limit is reached, or the iteration budget is exhausted; the optimal terminal node is 
returned as the investigation result.


The supervisor coordinates two agents through a reflection-triggered handoff protocol. Following the log agent's search completion,  the supervisor initiates handoff to the metric agent when the terminal reflection score $r < 0.7$ or diagnostic completeness $c_{\text{comp}} < 0.6$. The supervisor extracts four attributes from the preceding agent: (A1) a concise natural-language description of the key investigative findings, (A2) a normalized reflection score $r \in [0.0, 1.0]$, (A3) the evidence item count for that agent, and (A4) a boolean escalation flag indicating whether handoff was triggered. Complete search trees, including intermediate nodes, UCT values, action signatures, and tool invocation details, remain encapsulated within each agent and are never exposed to the supervisor or peer agents.  This privacy-preserving design prevents circular reasoning contamination. The metric agent is then invoked with an augmented query:
\begin{equation}
q_{\text{metrics}} = q_{\text{original}} \oplus s_{\text{log}},
\label{eq:handoff}
\end{equation}
where $\oplus$ denotes the augmentation of the original query with the log agent's investigative summary $s_{\text{log}}$ extracted in attribute~(A1). This formulation treats handoff as information transfer at the level of condensed evidence, ensuring that the receiving agent conditions on salient findings without inheriting the upstream search trajectory. This design reduces the risk of circular reasoning while enabling cross-modal validation.

After both agents complete their searches, the supervisor performs cross-modal correlation by prompting the LLM to compare both investigative summaries against their associated reflection scores and classify the level of agreement into one of four categorical labels: \emph{strong correlation} (metrics confirm the log-derived diagnosis), \emph{weak correlation} (partial overlap without clear causal alignment), \emph{no correlation} (metrics provide no evidence relevant to the log-derived diagnosis), or \emph{contradictory} (metrics contradict the log-derived diagnosis). The supervisor then generates the final diagnosis conditioned on both investigative summaries and the assigned correlation label.  


\section{Evaluation}
We defined the following research questions: 
\begin{itemize}
    \item \textbf{RQ1. Diagnostic accuracy}:  How does LATS-RCA perform in terms of root cause diagnostic accuracy?
    \item \textbf{RQ2. Computational cost}:  What computational cost is associated with achieving this diagnostic accuracy?
    \end{itemize}
\subsection{Experiment setup}
\subsubsection{Dataset.} We evaluate LATS-RCA on the LO2 publicly available dataset~\cite{L02paper}. LO2 is an open-source implementation of the OAuth2.0 authorization protocol. It consists of seven microservices and a MySQL database. The LO2 dataset is generated by executing the LO2 system under controlled API-level error injection. For each run, the system is first exercised with valid API requests, and then a specific API error is injected by issuing an erroneous request. During each execution, logs are collected for each service, while 485 performance metrics (e.g., CPU, memory, disk usage) are recorded at the system level via Prometheus; each run is labeled according to the injected API error type. This process is repeated across a large number of runs to construct a labeled dataset of normal and anomalous executions. In this study, we evaluate LATS-RCA using the lo2-sample dataset from LO2. We use this dataset as provided, without additional sampling or modification.  This dataset contains 100 sampled anomaly cases spanning 53 distinct failure types, including OAuth 2.0 protocol violations, CRUD failures, and authentication errors.



\subsubsection{Baselines.} Deep learning–based graph and causal approaches are not included in the comparison, as they primarily focus on fault localization by ranking suspicious components rather than identifying the underlying root cause, and therefore correspond to a different task formulation. Directly comparable LLM-based agent RCA baselines for MSS are not available under our experimental setting. We model RCA as a finite-horizon sequential decision process for label-based diagnosis, and operate on logs and metrics as the monitoring modalities. RCAgent and MicroRCA are the closest related approaches. However, in addition to logs and metrics, both approaches rely on distributed traces as the primary source of diagnostic evidence, which fundamentally differs from our setting. Moreover, they output structured diagnostic reports rather than a root cause, making their evaluation protocols incompatible with ours. Therefore, we construct two LLM-agent baselines within our experimental setting to compare LATS-RCA against the applicable approaches. Specifically, we compare LATS-RCA against two ReAct~\cite{yao2023react} baselines: (i) a single-agent ReAct baseline that uses the same tools but produces a linear reasoning trajectory, and (ii) a multi-agent ReAct variant that executes sequential log $\rightarrow$ metrics analysis. ReAct represents a widely adopted LLM-agent paradigm that follows a single-pass linear reasoning trajectory (similar to RCAgent and MicroRCA), where the agent alternates between diagnostic reasoning and tool invocation to refine its root cause hypothesis.



\subsubsection{Evaluation metrics.} We evaluate diagnostic effectiveness using accuracy, measuring whether the predicted root cause matches the ground-truth failure label for each scenario. In our setting, each input corresponds to a pre-identified anomalous scenario, and the task is to diagnose its root cause rather than to detect anomalies. Since each scenario in the LO2 dataset is associated with a single ground-truth root cause, the problem is formulated as a single-label diagnosis task, for which accuracy provides a direct and appropriate evaluation metric \cite{YuqingFSE2025}. Metrics such as precision, recall, and F1-score are commonly used in anomaly detection settings, but are not applicable here, as each instance requires exactly one prediction rather than a set of positive/negative decisions \cite{Varun2009}. In addition, we measure computational costs by the number of API calls, token consumption, and wall-clock runtime, following prior studies on of LLM reasoning evaluation \cite{han2025token}. We also analyze search behavior using the number of hypotheses explored and the number of evidence items collected, which capture the breadth of exploration and the extent of evidence usage during diagnosis. The number of hypotheses is counted as the total number of child nodes expanded during tree search, while evidence items are counted as the number of \texttt{ToolMessage} entries from agents, as defined in Section \ref{sec:LATS-RCA}.




\subsubsection{Implementation details.} We implement LATS-RCA in Python 3.12 using LangChain~\cite{langchain} for tool integration and LangGraph~\cite{langgraph} for agent orchestration. All agents (log, metric, and supervisor) use Claude Sonnet 4.5 
accessed via the Anthropic API (SDK \texttt{anthropic>=0.28,<0.30}), with temperature fixed at $\tau = 0.7$ for all LLM invocations, including policy generation, reflection scoring, and supervisor correlation. No additional sampling parameters are modified from SDK defaults. Since the Anthropic API does not support random seeds, stochasticity is isolated to LLM text generation; tool execution and all downstream computations (reward computation, UCT selection, and backpropagation) are deterministic. Experiments requiring variance estimation aggregate multiple independent runs. Additional implementation details are provided in our replication package. Experiments run single-threaded on commodity hardware with runtime dominated by API latency; wall-clock measurements are reported for transparency but excluded from primary performance metrics. 



\subsection{Study results}


\subsubsection{RQ1. Diagnostic Accuracy}
As shown in Table \ref{tab:lo2-main-results}, LATS-RCA achieves a diagnostic accuracy of 91.3\%. To account for stochasticity in LLM-based reasoning, we repeated the experiments across five runs and observed consistent performance, with a standard deviation of 1.2\%. LATS-RCA outperforms the single-agent ReAct baseline (39.8\%) and multi-agent ReAct baseline (57.4\%). This corresponds to improvements of about 50\% and 34 \%, respectively. 
The improvement over the single-agent ReAct baseline suggests that systematic exploration of multiple root-cause hypotheses is critical for accurate diagnosis under ambiguous evidence. The improvement over the multi-agent ReAct baseline indicates that multi-agent specialization and cross-modal handoff alone do not fully contribute to the gains; rather, the structured search process further strengthens diagnostic accuracy beyond linear cross-modal reasoning under the same tool and data setting. 


\begin{table}[H]
\centering
\caption{Evaluation results on the LO2 dataset.}
\label{tab:lo2-main-results}
\begin{tabular*}{\linewidth}{@{\extracolsep{\fill}}lrrrr}
\toprule
Method & Acc. (\%) & Calls & Tokens (K) & Time (min) \\
\midrule
LATS-RCA & 91.3 & 53.1 & 156 & 9.1 \\
ReAct (single) & 39.8 & 13.5 & 41 & 1.2 \\
ReAct (multi) & 57.4 & 25.2 & 73 & 1.7 \\
\bottomrule
\end{tabular*}
\end{table}

\subsubsection{RQ2. Computational Cost}
LATS-RCA incurs higher computational costs than the baseline approaches, see Table \ref{tab:lo2-main-results}. LATS-RCA requires 53.1 API calls per investigation compared to 13.5 for single-agent ReAct baseline and 25.2 for multi-agent ReAct baseline. Token consumption exhibits a similar trend: LATS-RCA uses 156K tokens, whereas the single-agent and multi-agent ReAct baselines consume 41K and 73K tokens, respectively. The investigation time is also increased: LATS-RCA takes 9.1 minutes per investigation, compared to 1.2 minutes for single-agent ReAct baseline and 1.7 minutes for the multi-agent ReAct baseline. 

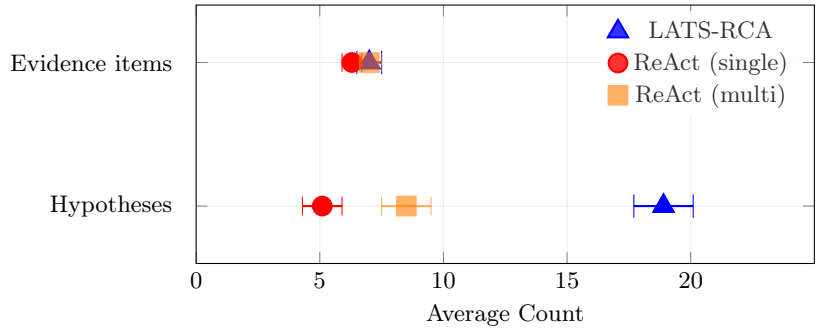
\begin{figure}[!htbp]
    \centering
    \begin{tikzpicture}[font=\small]
        \begin{axis}[
            width=0.8\columnwidth, 
            height=5cm,
            xlabel={Average Count},
            ytick={1,2},
            yticklabels={Hypotheses, Evidence items},
            yticklabel style={font=\footnotesize, xshift=-2mm}, 
            xmin=0, xmax=25,
            legend style={
                at={(0.98,0.98)},
                anchor=north east,
                font=\footnotesize,
                draw=none,
                fill=white,
                fill opacity=0.8
            },
            grid=major,
            grid style={line width=0.1pt, draw=gray!15},
            xlabel style={font=\footnotesize},
            xtick={0,5,10,15,20},
            enlarge y limits=0.4,
            mark options={line width=0.8pt}
        ]

        \addplot[
            only marks,
            mark=triangle*,
            mark size=4.4pt,
            blue,
            error bars/.cd,
            x dir=both,
            x explicit,
            error bar style={blue, line width=0.8pt}
        ] coordinates {
            (18.9,1) +- (1.2,0)
            (7.0,2) +- (0.5,0)
        };
        \addlegendentry{LATS-RCA}

        \addplot[
            only marks,
            mark=*,
            mark size=3.5pt,
            red,
            error bars/.cd,
            x dir=both,
            x explicit,
            error bar style={red, line width=0.8pt}
        ] coordinates {
            (5.1,1) +- (0.8,0)
            (6.3,2) +- (0.4,0)
        };
        \addlegendentry{ReAct (single)}

        \addplot[
            only marks,
            mark=square*,
            mark size=3.5pt,
            orange,
            opacity=0.6,
            error bars/.cd,
            x dir=both,
            x explicit,
            error bar style={orange, line width=0.8pt}
        ] coordinates {
            (8.5,1) +- (1.0,0)
            (7.0,2) +- (0.5,0)
        };
        \addlegendentry{ReAct (multi)}

        \end{axis}
    \end{tikzpicture}
    \vspace{-0.2cm}
    \caption{\textbf{Search behavior}: Dot plot comparing exploration breadth (hypotheses) versus depth (evidence). 
    }
    \label{fig:exploration-effectiveness}
    \vspace{-1em}
\end{figure}

The higher computational cost of LATS-RCA stems primarily from its broader hypothesis exploration rather than from increased evidence collection. As illustrated in Figure \ref{fig:exploration-effectiveness}, LATS-RCA explores an average of 18.9 hypotheses per investigation, compared to 5.1 for the single-agent ReAct baseline and 8.5 for the multi-agent ReAct baseline, representing a 3.7× increase in search breadth. At the same time, all methods collect comparable numbers of evidence items: approximately 7.0 selected evidence items per investigation. These results indicate that the diagnostic accuracy gains of LATS-RCA stem from systematic exploration and reflection-guided evaluation, rather than from gathering more monitoring data. By examining a larger set of alternative explanations while operating on the same underlying evidence items, LATS-RCA demonstrates more effective utilization of available observability signals.

\textit{Cost-accuracy trade-off}. LATS-RCA uses approximately 3.8× and 2.1× the tokens, and 7.6× and 5.4× the runtime, of single- and multi-agent ReAct, respectively. In production incident response, where the cost of a missed or delayed root cause (prolonged outage, SLA penalties) typically dwarfs API expenditure, this additional reasoning cost represents a favorable trade-off.

\subsection{Ablation Study}
We conduct an ablation study to understand the contribution of individual components in LATS-RCA. Table \ref{tab:ablation} reports the results. Overall, candidate batching contributes the most to performance, followed by backpropagation and reflection, and all three components are necessary to achieve 91.3\% diagnostic accuracy of LATS-RCA. Removing candidate batching (w/o candidate batching) leads to the largest performance degradation, with diagnostic accuracy dropping from 91.3\% to 84.3\%, indicating that maintaining multiple candidate hypotheses is critical for effective diagnosis. Disabling backpropagation (w/o backpropagation) results in the second-largest drop to 84.8\%, suggesting that propagating feedback from successful reasoning paths plays an important role in guiding the search process. In contrast, removing the reflection mechanism (w/o reflection) causes a smaller but still noticeable decrease to 87.6\%, showing that reflection contributes less than the other components but remains beneficial. 

\begin{table}[h]
\centering
\caption{Ablation study.}
\label{tab:ablation}
\begin{tabular*}{\textwidth}{@{\extracolsep{\fill}}rrr}
\toprule
Variant & Acc. (\%) & $\Delta$ (pp) \\
\midrule
LATS-RCA & 91.3 & 0.0 \\
w/o candidate batching & 84.3 & -7.0 \\
w/o backpropagation & 84.8 & -6.5 \\
w/o reflection & 87.6 & -3.7 \\
\bottomrule
\end{tabular*}
\end{table}
\vspace{-0.5cm}

We further evaluate LATS-RCA's sensitivity to LLM backbone choice (Table~\ref{tab:backbone}). Accuracy varies by at most 1.6 \% across frontier-tier models (Claude Sonnet 4.5, GPT-5, Gemini 3 Pro). These models also show similar computational cost per investigation. 
Simpler backbones require more API calls and tokens (e.g., GPT-5.2 nano: 68.2 calls and 201K tokens vs. Sonnet 4.5: 53.1 calls and 156K tokens) yet achieve lower accuracy, suggesting that increased search may not fully compensate for weaker reasoning capacity.


\begin{table}[ht]
\centering
\caption{LATS-RCA performance across LLM backbones on LO2}
\label{tab:backbone}
\begin{tabular}{lrrrrr}
\toprule
Model & Acc.(\%) & Calls & Tokens (K) & Time (min) \\
\midrule
Claude Sonnet 4.5 & 91.3 & 53.1 & 156 & 9.1  \\
GPT-5             & 89.7  & 56.3 & 168 & 10.4  \\
Gemini 3 Pro      & 90.1  & 54.8 & 171 & 8.7  \\
Claude Haiku 4.5  & 79.6  & 61.4 & 183 & 4.2  \\
GPT-5.2 nano      & 72.8 & 68.2 & 201 & 5.8  \\
\bottomrule
\end{tabular}
\end{table}
\vspace{-1cm}

\section{Exploratory Study in a Production Environment}
\subsubsection{Production Environment.} 
We deployed LATS-RCA in Zoner Oy's real production MSS (Prod) that consists of seven core services implemented in a polyglot stack: Node.js for API gateway and orchestration layers, Rust for hosting services managing high-performance WordPress infrastructure, Python for data processing pipelines, and Go for infrastructure services. The architecture follows a hybrid communication model using NATS for event-driven messaging, gRPC for synchronous service-to-service calls, and HTTP/REST for external API endpoints. 
Prod runs on a self-hosted Kubernetes cluster as part of an internal developer platform, following cloud-native patterns with containerized services, horizontal pod auto-scaling, and service mesh observability. Production workloads serve over 300,000 websites across Europe, including more than 5,000 high-performance managed WordPress hosting customers in Finland. 
Observability infrastructure captures logs through container stdout/stderr aggregation, generating approximately 8,000 to 25,000 log entries per service per hour during peak traffic. Metrics are collected via Prometheus exporters at 15-second intervals, tracking service-level indicators (request rates, latencies, error rates) and infrastructure metrics (CPU, memory, network I/O).

\subsubsection{Data Characteristics and Challenges.}

We collected 37 real-world incidents over a six-month period from the deployed production system and constructed a dataset from their corresponding logs and metrics. We manually analyzed each incident and independently verified its root cause(s), ensuring reliable ground truth. These incidents span five failure categories: configuration errors (11), resource exhaustion (9), upstream dependency failures (8), deployment regressions (5), and infrastructure faults (4). Each incident encompasses 53,000–217,000 log lines and 300–800 metric time series within 15-45 min
observation windows, reflecting the scale and complexity of real-world production observability data. 

The data exhibits substantial heterogeneity in both logs and metrics across services. First, due to the polyglot service stack, logs appear in multiple formats, including JSON-structured entries from Node.js and Python services, unstructured text from Go, and mixed formats from Rust. Logs further vary in timestamp formats (e.g., ISO 8601 with timezone offsets, epoch milliseconds, and architecture-specific patterns), severity taxonomies across logging frameworks (e.g., Java’s SEVERE/WARNING, Python’s CRITICAL/ERROR), and structural conventions such as JSON records, key–value pairs, and unstructured narratives. Second, production metrics exhibit similar variability. They originate from diverse Prometheus exporters, with service-specific naming conventions, varying units, and partial coverage across incidents. Such heterogeneity in logs and metrics introduces normalization and alignment challenges that are not present in the LO2 dataset.

To address data heterogeneity, we apply a systematic normalization pipeline to them before applying LATS-RCA. For logs, we convert heterogeneous timestamp formats to a unified UTC ISO 8601 representation with millisecond precision, map severity levels to LO2’s canonical hierarchy, extract key diagnostic fields (e.g., trace id, service name, and error code) into uniform metadata prefixes, and aggregate multiline stack traces into coherent log entries. For metrics, we align production Prometheus instrumentation with LO2’s runtime schema through naming-pattern matching and unit conversion, and mark unavailable metrics rather than synthesizing them. We verify the normalization process using both automated consistency checks and targeted manual inspection, ensuring that diagnostic-critical information required for root cause analysis is preserved.

\subsubsection{Production validation.} Due to the stochastic nature of LLMs (e.g., temperature and sampling), we repeat each experiment multiple times. Across runs, LATS-RCA correctly diagnoses an average of 24.1 out of 37 production incidents (65.1\%), with results ranging from 62\% to 70\%. This is lower than the 91.3\% accuracy observed on LO2, where performance remains stable across runs with a standard deviation of 1.2\%. The computational cost increases substantially in this production setting due to data scale and heterogeneity, requiring more reasoning steps. On average, each incident involves approximately 75 API calls, 220K tokens, and 13 minutes of execution time. We observe that the gap in diagnostic accuracy and reasoning cost between LO2 and Prod can be attributed to several challenges of real-world RCA, as discussed below.

\textit{Multi-factor root causes}. In our production system, incidents often involve multiple interacting factors. This complicates both diagnosis and evaluation: identifying one of two interacting causes constitutes partial correctness that binary evaluation cannot adequately capture. In contrast, benchmark datasets such as LO2 and other widely used MSS RCA datasets (e.g., Nezha \cite{Nezha2023}, DeepTraLog \cite{DeepTraLog}, and AIOps challenge datasets \cite{li2022constructinglargescalerealworldbenchmark}) rely on injected faults with a single ground-truth root cause for controlled evaluation. This simplifying assumption, while enabling controlled evaluation, does not fully reflect the complexity of real-world production failures. This observation aligns with prior empirical studies on microservice RCA, which note that benchmark settings often simplify failure conditions and may not fully reflect real-world complexity \cite{fang2025rethink,RCALuan2024}.

\textit{Production scale and dependency complexity}. Our production system exhibits substantially larger service dependency graphs and more complex inter-service dependencies than those typically found in widely-used controlled benchmark settings (e.g., LO2, Nezha, DeepTraLog, or AIOps challenge datasets). This increases the number of possible root-cause explanations, making diagnosis more challenging and requiring deeper reasoning. This will contribute to both the lower diagnostic accuracy and the higher reasoning cost observed in production. 

\textit{Incomplete observability}. In our production system, metrics instrumentation is often incomplete or inconsistent, resulting in partial evidence and increased uncertainty for LLM-based diagnostic reasoning. This challenge is further complicated by environmental variability, such as undocumented service dependencies, transient infrastructure faults, and unpredictable user behavior, which are typically absent from widely-used controlled benchmark settings.

Despite these challenges, our results show that LATS-RCA’s reflection-guided tree search remains applicable in operational environments. Its deployment also reveals three challenges of real-world incident RCA, including multi-factor root causes, scale and dependency complexity, and incomplete observability, which are largely absent from controlled benchmarks. These findings suggest promising directions for future RCA research, including support for multi-factor root cause reasoning, robustness under incomplete observability, and scalability to large heterogeneous service environments.

\section{Threats to validity}
External validity may be affected by the gap between benchmark datasets and real-world production environments. RCA benchmarks are typically designed with injected faults and clear single root causes to enable controlled and reproducible evaluation, but this simplification does not capture key characteristics of production systems, such as multi-factor failures, heterogeneous observability data, and polyglot technology stacks. To mitigate this limitation, we use the LO2 dataset, which covers a broader range of fault types than simpler benchmarks, and further validate LATS-RCA in a large-scale production system. The differences observed between LO2 and production are largely attributable to the mismatch between controlled benchmark settings and real-world contexts.

There are construct validity considerations related to the evaluation methodology. In real-world complex MSS, anomalies may involve multiple contributing root causes, and diagnostic outcomes are not always binary. For experimental evaluation, this study follows the dataset-defined single-label notion of root cause and assesses correctness using an exact-match criterion. This definition is aligned with the fault-injection design of the benchmark and supports reproducible comparison across approaches under a shared evaluation setting.



\section{Conclusion}

This paper presents LATS-RCA, a Language Agent Tree Search for RCA in MSS. By formulating RCA as a reflection-guided tree-structured search over logs and metrics at the microservice level, LATS-RCA goes beyond single-path linear reasoning. We evaluate LATS-RCA on the LO2 benchmark, where it achieves high diagnostic accuracy under controlled settings. We further conduct an exploratory study by validating LATS-RCA in the Prod. Compared to LO2, performance in Prod shows lower diagnostic accuracy and higher reasoning cost. In our production study, we observe three challenges of real-world RCA that are not adequately captured in benchmark settings, including multi-factor root causes, large-scale system complexity, and incomplete observability. These factors contribute to the observed gap in diagnostic accuracy and reasoning cost between benchmark and production settings. Despite these challenges, the results demonstrate the practical applicability of LATS-RCA in real-world environments. Future work will focus on addressing the challenges observed in production settings, including improving support for multi-factor root cause reasoning, enhancing robustness under incomplete observability, and scaling diagnostic methods to handle large and complex service dependencies. Additionally, new benchmark datasets are needed that include novel anomalies, such as multi-factor root causes, as observed in our industrial data.

\section{Disclosure of Interests}
The authors have no competing interests to declare that are relevant to the content of this article.

\section{Acknowledgment}
This work is funded by the EuroHPC Joint Undertaking and its members including top-up funding by the Ministry of Education and Culture. This work is also supported by the Research Council of Finland (grant id: 359861, the MuFAno project).



\bibliographystyle{splncs04}
\bibliography{mainbib}
\end{document}